\documentstyle[twocolumn,prl,aps,epsf]{revtex}
\begin{document}

\draft
\wideabs{
\title{Spin Fluctuation Induced Dephasing in a Mesoscopic Ring}
\author{Kicheon Kang$^1$, Sam Young Cho$^2$, and Kyoung Wan Park$^1$}
\address{   $^1$Basic Research Laboratory, Electronics and Telecommunications
            Research Institute, Taejon 305-350, Korea}
\address{   $^2$Department of Physics, University of Queensland, Brisbane 
            4072, Australia }

\date{\today}

\maketitle

\begin{abstract}
We investigate the persistent current in a hybrid Aharonov-Bohm
ring - quantum dot system coupled to a reservoir which provides 
spin fluctuations.
It is shown that the spin exchange interaction between the quantum dot
and the reservoir induces dephasing in the absence of direct charge transfer.
We demonstrate an anomalous nature of this spin-fluctuation induced dephasing
which tends to {\em enhance} the persistent current. We explain our result
in terms of the separation of the spin from the charge degree of freedom. 
The nature of the spin fluctuation induced dephasing is analyzed in detail. 
\end{abstract}
}
%
%

Persistent current (PC) in a mesoscopic Aharonov-Bohm (AB) ring is an ideal 
probe of quantum coherence of electron motion in the equilibrium state. 
Usually the PC is likely to be suppressed by 
various dephasing processes. The role of the intrinsic dephasing at low
temperature has not
been well understood till now~\cite{mohanty97}.
An alternative viewpoint to this is to introduce
an artificial dephasor, in order to study the effect of 
decoherence in a controlled manner~\cite{buks98}. 
A conceptually simple but instructive example for
that purpose is an AB ring attached to an electron reservoir which 
exchanges charges with the ring~\cite{buttikker85}. 
In the reservoir electrons are scattered
inelastically and there is no phase coherence between electrons absorbed
and those emitted by the reservoir. Therefore charge
transfer between the ring and the reservoir diminishes the coherence and 
thus the AB oscillation. 
On the other hand, the effect of the spin exchange interactions 
on the PC has been attracting growing interest in recent
years~\cite{ferrari99,kang00,eckle01,cho01,affleck01,hu01,anda01}.
It has been proposed that the spin fluctuation affects the PC in
a drastically different manner compared to the case of charge
fluctuation~\cite{kang00,eckle01,cho01,kang01}.
Experimentally, the role of coherent spin fluctuation has been investigated
by transport measurements using AB interferometer 
setup~\cite{wiel00,ji00}

In this Letter, we address the effect of dephasing induced
by the spin fluctuations. For this purpose
we consider the geometry schematically drawn in Figure 1, where the spin
fluctuation between the ring and the reservoir is mediated via 
antiferromagnetic exchange
interactions with the quantum dot (QD), while direct charge transfer is
prohibited by the Coulomb blockade. We find a counterintuitive result
that the dephasing tends to {\em enhance} the PC rather 
than to reduce it in this geometry. We argue that this enhancement 
can be regarded as a signature
of the separation of the spin degree of freedom from the charge one.

Our model is described by the Hamiltonian:
\begin{equation}
\label{eq:model}
 H = H_0 + H_R + T \;,
\end{equation}
where $H_0$, $H_R$ and $T$ stand for the hybrid dot-ring system, 
reservoir, and tunneling between the QD and the reservoir, respectively.
$H_0$ is decomposed into the three parts as
\begin{equation}
 H_0 = H_{QD} + H_{ring} + H_{t'} \;,
\end{equation}
where $H_{QD}$, $H_{ring}$, and $H_{t'}$
correspond to the Hamiltonians describing
the quantum dot, the AB ring, and the dot-ring hybridization, respectively:
\begin{mathletters}
\begin{eqnarray}
 H_{QD} &=&  \sum_\sigma \varepsilon_d d_\sigma^\dagger d_\sigma
       + U {n}_\uparrow {n}_\downarrow \;, \\
 H_{ring} &=& -t \sum_{j=1}^{N}\sum_\sigma
     \left(  e^{i\varphi/N} c^\dagger_{j\sigma}
           c_{j+1\sigma} + \mbox{\rm h.c.}\right) \label{eq:hring} \;, \\
 H_{t'} &=& -t' \sum_\sigma \left( d^\dagger_\sigma
                c_{0\sigma} + c_{0\sigma}^\dagger d_\sigma \right) .
\end{eqnarray}
\end{mathletters}
Here, we describe the ring by using a tight-binding Hamiltonian with 
$N$ lattice
sites, and the QD by a single Anderson impurity.  The single particle
energy and the on-site Coulomb repulsion in the QD are represented by
$\varepsilon_d$ and $U$, respectively.
The phase $\varphi$ in Eq.~(\ref{eq:hring}) comes from the AB
flux, and is defined by $\varphi = 2\pi\Phi/\Phi_0$, where $\Phi$ and
$\Phi_0$ are the external flux and the flux quantum ($=hc/e$), respectively.
Note that Eq.~(\ref{eq:hring}) can be diagonalized and the
corresponding eigenvalues are given by
$\varepsilon_m = -2t\cos{\frac{1}{N} (2\pi m - \varphi)}$ ($m$ being an
integer number). The reservoir is modeled by 
a Fermi sea of electrons with single
particle energies \{$E_k$\} :
\begin{equation}
 H_R = \sum_{k\sigma} E_k a_{k\sigma}^\dagger a_{k\sigma} \; ,
\end{equation}
Finally, coupling between the QD and the reservoir is
given by a tunneling Hamiltonian
\begin{equation}
 T = \sum_{k\sigma} \tau_{k}
 \left( a_{k\sigma}^\dagger d_\sigma + \mbox{ h.c.} \right) \;.
\end{equation}

The hopping strength of reservoir-dot and that of ring-dot 
are represented by $\Gamma_R$ and $\Gamma'$, respectively.
$\Gamma_R$, assumed to be constant at the energy interval of $-D<\varepsilon
<D$, is defined as
\begin{equation}
 \Gamma_R = \pi \rho_R(\varepsilon) |\tau(\varepsilon)|^2  \;,
\end{equation}
where $\rho_R$ and $\tau$ represent the density of states in the reservoir
and the tunneling amplitude between the reservoir and the QD.
For half-filled case (where the Fermi energy is set to be zero) in the
continuum limit, $\Gamma'$ can be simply written as~\cite{kang01}
\begin{equation}
 \Gamma' = \frac{t'^2}{2t} \;,
\end{equation}
and the level discreteness at the Fermi energy is given by
\begin{equation}
 \delta = \frac{2\pi t}{N} \;.
\end{equation}
Our study is restricted to the simplest half-filled case, which does not
affect the result and conclusion we will draw.

To rule out the effect of charge fluctuation, we consider the parameter 
limit of $-\varepsilon_d, \varepsilon_d+U \gg \Gamma'+\Gamma_R$. 
In the absence of the reservoir, the Bethe ansatz result~\cite{eckle01}
shows that the PC is not affected by the QD in the
Kondo limit with $\delta/T_K\rightarrow0$, 
where $T_K$ stands for the Kondo energy scale.
For finite value of $\delta/T_K$, the coupling of the ring to the QD
linearly reduces the PC as a function of $\delta/T_K$
for small $\delta/T_K$, and induces a crossover from the
continuum Kondo limit ($\delta/T_K\ll 1$) to an effectively decoupled
ring-dot system ($\delta/T_K\gg 1$)~\cite{cho01}.

To calculate the PC in the presence of the reservoir, we adopt the leading
order $1/N_s$-expansion with $N_s$ being the magnetic degeneracy.
This approach was shown to describe well the essence of the Kondo
correlation preserving the Fermi liquid properties~\cite{hewson93}.
In addition, for the ring-dot system without reservoir, this approximation
reproduces the exact Bethe ansatz result for 
$\delta/T_K\rightarrow0$~\cite{cho01}.
Here we consider the infinite-$U$ limit since the consideration
of finite $U$ and double occupancy in the QD does not provide any 
modification
to the renormalized physical quantities~\cite{kang96}.
Then the problem is reduced to solving 
the self-consistent equation:
\begin{equation}
 E_G' = \frac{\Gamma'}{\pi}\delta  
  \sum_{m\sigma} \frac{f(\varepsilon_m) }{ 
  E_G'+\varepsilon_m-\varepsilon_d} 
 + \sum_{k\sigma}  \frac{f(E_k)|\tau_k|^2 }{ E_G'+E_k-\varepsilon_d} 
   \; , \label{eq:energy}
\end{equation}
where $f(\varepsilon)$ is the Fermi distribution function.
$E_G'\equiv E_G-E_\Omega$, where $E_\Omega$ is the energy
of the ground state without tunneling. $E_G$
corresponds to the ground state energy of the coupled system
at zero temperature.

At zero temperature the 2nd term of Eq.~(\ref{eq:energy}) can be
calculated analytically and the latter is rewritten as
\begin{equation}
 E_G' = 2\frac{\Gamma'}{\pi}\delta  
  \sum_{\varepsilon_m<0} \frac{ 1 }{ E_G'+\varepsilon_m-\varepsilon_d}
 + \frac{2\Gamma_R}{\pi} \log{\frac{\varepsilon_d-E_G'}{ D } } \;.
  \label{eq:energy0}
\end{equation}
The PC is defined in terms of the phase sensitivity of the
ground state energy as
\begin{equation}
 I(\varphi) = -\frac{e}{\hbar} \frac{\partial E_G}{\partial\varphi} \;.
  \label{eq:pc}
\end{equation}
Combining Eq.(\ref{eq:pc}) with Eq.(\ref{eq:energy0}),
the PC can be expressed as the sum of two terms, 
$I_{ring}$ and $I_{int}$, originating from
the ideal ring and from the interactions, respectively:
\begin{mathletters}
\begin{eqnarray}
 I(\varphi) &=& I_{ring}(\varphi) + I_{int}(\varphi) \;, \\
 I_{ring}(\varphi) &=& 2\sum_{\varepsilon_m<0} I_m(\varphi) \;, \\
 I_{int}(\varphi) &=& -2\frac{\Gamma'}{\pi} \delta {\cal Z}
   \sum_{\varepsilon_m<0}
   \frac{1}{ E_G'+\varepsilon_m-\varepsilon_d }  I_m(\varphi) \;, 
  \label{eq:i_qd}
\end{eqnarray}
where 
 \begin{equation}
 I_m(\varphi) = -\frac{e}{\hbar}\frac{\partial\varepsilon_m}{\partial\varphi}
 \end{equation}
is the contribution to the PC from the bare ring energy level $\varepsilon_m$,
and 
\begin{equation}
 {\cal Z} = \left( 1 + 2\frac{\Gamma'}{\pi} \delta 
   \sum_{\varepsilon_m<0} \frac{1}{( E_G'+\varepsilon_m-\varepsilon_d)^2}
 + \frac{2\Gamma_R}{\pi}\frac{1}{ \varepsilon_d-E_G' }
 \right)^{-1}
\end{equation}
corresponds to the renormalization constant for the ground state.
\end{mathletters}
Note the negative sign at the R.H.S. of Eq.(\ref{eq:i_qd}) which leads
to the reduction of the PC for finite value of $\delta/T_K$. 

Figure 2 displays the effect of the coupling to the reservoir on the
PC.  The PC is plotted as a function of the dimensionless coupling strength
$\gamma$ defined by 
\begin{equation}
 \gamma \equiv \Gamma_R/\Gamma' \;.  
\end{equation}
Note that the PC is a universal function of $\gamma$, $\varphi$,
and $\delta/T_K^0$,
which does not depend on the parameter detail chosen here.
$T_K^0$ in the figure denotes the Kondo energy scale in the absence 
of the coupling 
to the reservoir ($\Gamma_R=0$) for $\delta\rightarrow0$,
(while $T_K$ stands for the counterpart including the reservoir).
First, one should recall that for $\Gamma_R=0$ the PC is reduced 
to $I_{ring}$ in the continuum limit ($\delta\ll T_K^0$), and is suppressed
for finite value of $\delta/T_K^0$~\cite{cho01}. It will be natural
to believe
that the reservoir would reduce the AB oscillation since it is expected
to play the same role
as a charge reservoir which induces decoherence. However,
the result is {\em opposite} to this simple expectation. Coupling to the
reservoir {\em enhances} the PC as shown for several
values of $\delta/T_K^0$. As $\gamma$ increases, the PC is enhanced
and eventually it saturates to $I_{ring}$ for sufficiently large
$\gamma$.  

This anomalous result is interpreted as follows.
Instead of reducing the PC, coupling to reservoir suppresses $I_{int}$ only, 
the contribution originating from the spin exchange interactions. This
makes net increase of the PC because the direction $I_{int}$ is opposite
to $I_{ring}$. This is a unique signature that the spin and the charge degrees
of freedom are decoupled.  That is, the reservoir degrades
the coherence of spin degree of freedom ($I_{int}$) only,
while it does not affect the charge one ($I_{ring}$).

To be more precise, the influence of reservoir on the system can 
be classified into two factors.
(i) Increase of the Kondo binding energy: 
As the coupling turns on, the effective spin exchange interactions
between the electrons in the QD and the conduction electrons increase. 
This results in the 
enhancement of the binding energy (or reducing the size of the Kondo
screening cloud). (ii) Decoherence of electrons:
The Kondo scattering provides effective charge flow between the ring and
the reservoir. Since the electrons in the reservoir are scattered
inelastically, no phase coherence exists between the electrons absorbed
and those emitted by the reservoir. This feature has never been addressed
before in the Kondo limit.

Both effects ((i),(ii)) are present in the result of Figure 2. 
The effect of enhanced Kondo binding energy is not a
unique feature of our geometry. That is, the energy scale is modified
by changing other parameters as well, and it can be well understood in
terms of the renormalized energy scale, $T_K$.
In order to extract the dephasing
effect, we show the PC as a function of $\delta/T_K$ for 
three different values of $\gamma$ in Figure 3(a). The effect (i)
is already included in the renormalized parameter 
$\delta/T_K$ with $\gamma$-dependent $T_K$.
The PC displays a crossover at $\delta/T_K\sim 1$ from the Kondo limit 
($\delta/T_K\ll1$) to the effectively decoupled limit ($\delta/T_K\gg1$),
regardless of the coupling to the reservoir. For large $\delta/T_K$ the 
PC saturates to the value which corresponds to that of the ideal ring
with one electron subtracted from the present system (denoted by
the dotted line). This demonstrates effective decoupling of the ring
from the rest part of the system. 

Here we point out that the PC increases as $\gamma$ increases
even after subtracting the effect
of rescaled Kondo energy, as shown in Figure 3(a). 
This demonstrates the anomalous nature
of the Kondo-assisted dephasing that enhances AB oscillation. That is,
the dephasing influences only the interaction part of the current, $I_{int}$,
through the spin-fluctuation channel, which results in a net increase
of the PC. This property is analyzed in more detail in Figure 3(b).
For small $\delta/T_K$, $I_{int}/I_{ring}$ shows $\gamma$-dependent
linear behavior as $I_{int}/I_{ring} = -c(\gamma)\delta/T_K$, where
$c(\gamma)>0$. One can see that the slope $c(\gamma)$ decreases as $\gamma$
increases. The reduction of the slope as a function of $\gamma$ is
the result of dephasing through spin exchange interactions.

To complete our discussion it is instructive 
define the coherence factor, $\eta$, associated with the
spin-fluctuation-induced dephasing by the ratio
\begin{equation}
 \eta \equiv \frac{c(\gamma)}{c(0)} \;.
   \label{eq:c-factor}
\end{equation} 
The coherence factor (Figure 4) decreases monotonically
as $\gamma$ increases.
This is a manifestation of decoherence
mediated by the spin fluctuations.
This behavior is quite universal independent of any parameter
detail.

In conclusion, we have investigated the effect of reservoir coupled
to a composite AB ring - QD system on the PC. In the Coulomb blockade limit,
spin fluctuations induce decoherence of the system in an anomalous way.
The persistent current circulating the ring is enhanced due to the
dephasing in the Kondo limit. We have argued that this enhancement 
is closely related to the separation of the spin from the charge degree
of freedom. 

This work has been supported by the National Research Laboratory program
of the Korean Ministry of Science and Technology, and also by the
Korean Ministry of Information and Communication.


%
\begin{figure} 
\epsfxsize=3.0in
\epsffile{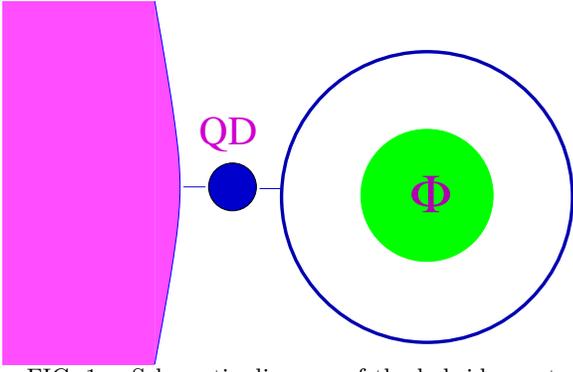}
 \caption{ Schematic diagram of the hybrid quantum dot - AB ring structure
 coupled to a reservoir. 
    }
  \label{fig1}
\end{figure}
\begin{figure} 
\epsfxsize=3.0in
\epsffile{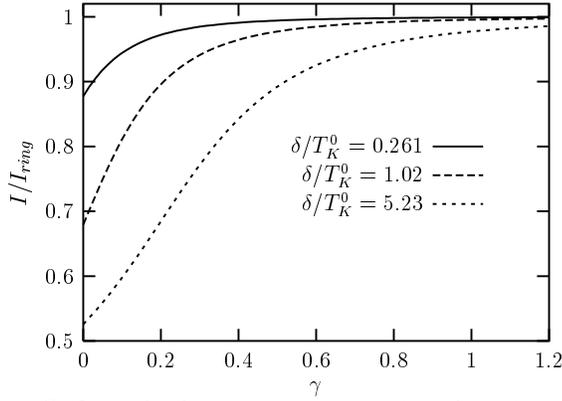}
 \caption{ Persistent current as a function of the renormalized coupling
 strength of the reservoir to the QD ($\gamma$) for $\Gamma'=0.125t$, 
 $\varepsilon_d=-0.7t$, and $\varphi=0.1\pi$ with several
 values of $\delta/T_K^0$. 
    }
  \label{fig2}
\end{figure}
\begin{figure}
\epsfxsize=3.0in
\epsffile{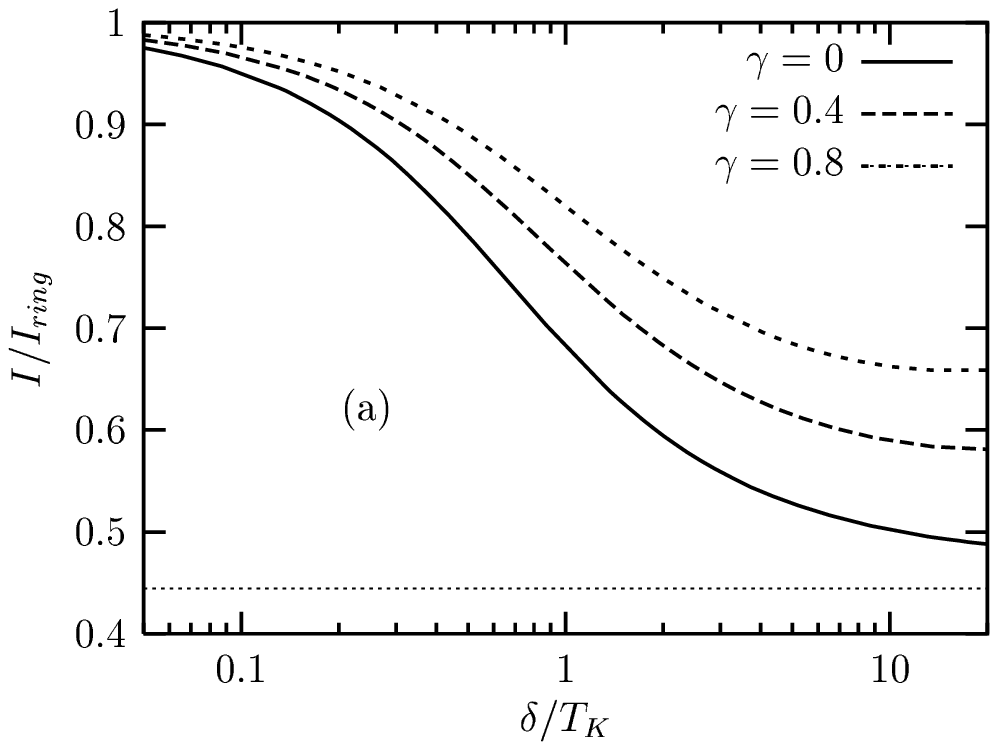}
\epsfxsize=3.0in
\epsffile{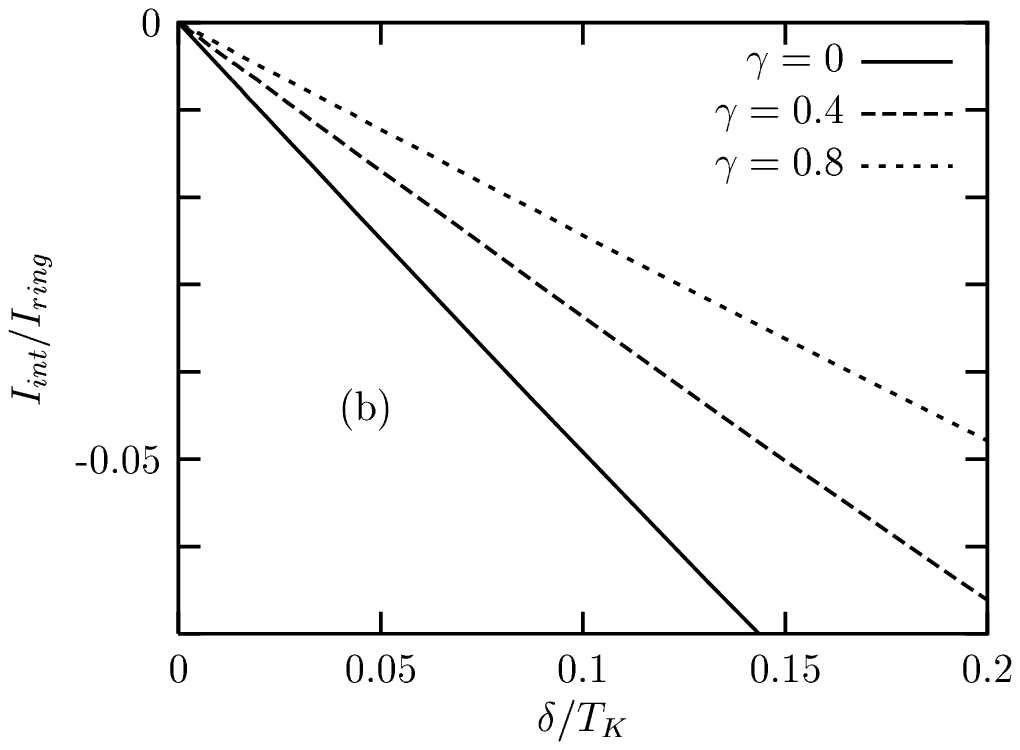}
 \caption{ (a) Persistent current as a function of the renormalized level
 spacing ($\delta/T_K$) for three different values of $\gamma$. Other
 parameters are given the same as those in Figure 2. 
 The dotted line indicates
 the PC of the ideal ring with one electron subtracted. (For $\varphi
 = 0.1\pi$, it corresponds to $\frac{4}{9} I_{ring}$.) 
 (b) Interaction contribution
 to the persistent current for the same parameters.
    }
  \label{fig3}
\end{figure}
\begin{figure}
\epsfxsize=3.0in
\epsffile{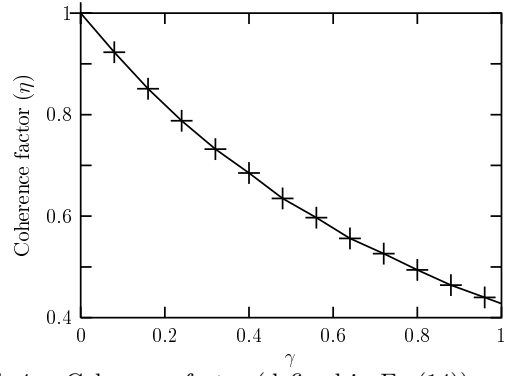}
 \caption{ Coherence factor (defined in Eq.(\ref{eq:c-factor})) 
 as a function of $\gamma$. 
    }
  \label{fig4}
\end{figure}
\end{document}